# Simulations and Design of a Single-Photon CMOS Imaging Pixel Using Multiple Non-Destructive Signal Sampling

**Konstantin D. Stefanov** [1,*], **Martin Prest** [1], **Mark Downing** [2], **Elizabeth George** [2], **Naidu Bezawada** [2] **and Andrew D. Holland** [1]

[1] Centre for Electronic Imaging, The Open University, Walton Hall, Milton Keynes MK7 6AA, UK
[2] European Southern Observatory, Karl-Schwarzschild-Strasse 2, D-85748 Garching, Germany
* Correspondence: Konstantin.Stefanov@open.ac.uk; Tel.: +44-1908-332116

**Abstract:** A single-photon CMOS image sensor design based on pinned photodiode (PPD) with multiple charge transfers and sampling is described. In the proposed pixel architecture, the photogenerated signal is sampled non-destructively multiple times and the results are averaged. Each signal measurement is statistically independent and by averaging the electronic readout noise is reduced to a level where single photons can be distinguished reliably. A pixel design using this method has been simulated in TCAD and several layouts have been generated for a 180 nm CMOS image sensor process. Using simulations, the noise performance of the pixel has been determined as a function of the number of samples, sense node capacitance, sampling rate, and transistor characteristics. The strengths and the limitations of the proposed design are discussed in detail, including the trade-off between noise performance and readout rate and the impact of charge transfer inefficiency. The projected performance of our first prototype device indicates that single-photon imaging is within reach and could enable ground-breaking performance in many scientific and industrial imaging applications.

## 1. Introduction

Single-photon (SP) imaging offers the ultimate performance in an imaging system due to its ability to capture and register each incoming photon [1][2]. It is particularly valuable in low light level conditions where every photon is precious, such as in astronomy, adaptive optics, night vision, surveillance and bio-imaging. In silicon, the most widely used semiconductor for image sensors, a single-photon with wavelength between 300 nm and 1100 nm can generate only one electron-hole pair. Therefore, for visible and near-infrared light the task of single-photon detection becomes a task of single electron (or hole) detection. This is not easy due to the unavoidable readout noise of the sensor, which is usually too high for a reliable single electron detection. Another difficulty for room temperature applications is the thermal dark current because it is indistinguishable from the photogenerated signal, however dark signal can be reduced to negligible levels by cooling.

Two main methods for achieving SP sensitivity are used in semiconductor image sensors. In the first one the photogenerated charge is amplified internally by a physical process before the conversion to voltage. In this way the signal is lifted well above the noise floor, allowing reliable SP detection. Typical examples are single-photon avalanche photodiodes (SPAD) and electron multiplying CCDs (EMCCD), in which the primary photogenerated electron undergoes avalanche multiplication. Both can resolve single photons, however they suffer from shortcomings including high dark current rate and after-pulsing in SPADs [3] and spurious charge and excess noise [4] in EMCCDs due to traps and the use of high voltages.

The second method involves reducing the readout noise of a sensor to a fraction of one electron RMS equivalent noise charge (ENC). Studies have shown [5] that for a practical single-photon imager with negligible error rate the ENC must be below 0.15 e- RMS. Recent advances in CMOS image sensor (CIS) technology have reduced the readout noise significantly, and CIS with ENC below 0.3 e- RMS have been reported [6]-[8]. These developments are due to the increase of the conversion gain of the sensors above 200 µV/e- by the use of special design and processing techniques, as well as by improvements to the noise performance of MOSFETs. Further noise improvements using those methods are certainly possible, however the difficulties increase as the noise approaches the required level of 0.15 e- RMS.

Recently, a new take on the well-known "skipper CCD" technique [9]-[11] has demonstrated deep sub-electron readout noise [12][13]. In skipper CCDs the charge is transferred under a floating gate in the buried channel (BC) multiple times and is measured after each transfer. Because the measurements are non-destructive and nearly statistically independent, averaging reduces the readout noise by the square root of the number of readouts. In [13] readout noise of only 0.068 e- RMS has been achieved after 4000 measurements, using an amplifier exhibiting 3.55 e- ENC in a single measurement. Similar method has been used for DEPFETs [14] to achieve 0.18 e- RMS noise in a sensor with inherent 3.1 e- ENC. The skipper technique is attractive because it can achieve sub-electron noise performance using designs, processes and MOSFETs available today. However, a major disadvantage of the multiple readouts is the greatly increased readout time, reaching several hours for CCDs [13]. For a more practical device the readout time must be reduced to at least a few seconds (e.g. for slow astronomical imaging), and to few milliseconds for applications requiring much higher frame rate.

CMOS image sensors could potentially offer the needed increase in speed due to the inherent parallelism in their readout. In the vast majority of CIS an entire row of pixels is read out simultaneously, which could reduce the readout time by three orders of magnitude compared to a sensor with a single or a small number of outputs such as the CCD. In addition, the finer feature size in modern CIS allows the conversion gain of a floating gate readout circuitry to be much higher than in CCDs. This finer feature size enables substantial reduction of the readout noise in a single measurement, necessitating fewer signal samples to be averaged.

Sub-electron readout noise CMOS image sensors capable of high readout rate would be useful for many applications. In the field of astronomy, large-format sensors with a size of at least 4K x 4K pixels, readout at a rate of a few frames a second would allow measurement of transient phenomena or observation techniques that require high time resolution and low noise, such as speckle imaging. Lucky imaging [15], for example, works best with noiseless image stacking and could benefit from this development. Smaller sensors that can be readout at rates of 0.5-1 kHz would replace EMCCDs in adaptive optics applications. Similarly, in bio-imaging applications such as molecular fluorescence imaging, these larger format sensors would enable microscope images covering a larger field with single-photon sensitivity.

In this paper we investigate the suitability of CIS-based skipper designs for achieving deep sub-electron readout noise for single-photon visible light imaging. The operating principles and TCAD simulations are presented in Section 2, followed by a noise analysis in Section 3, and a description of a proposed layout and its expected performance in Section 4.

## 2. Operating Principles

The most widely used pixel architecture in modern CIS is based on the pinned photodiode (PPD) [16]. The PPD achieves very low dark current due to the pinning implant, low image lag and high conversion gain by using a small sense node separate from the charge collection region. The PPD relies on efficient transfer of charge to the sense node, where the conversion to voltage occurs after the charge has been collected. Due to its excellent electro-optical characteristics and charge transfer capabilities, the PPD is one candidate for constructing a pixel capable of multiple signal sampling.

Another possibility is a low voltage CCD manufactured in a CMOS process, as used in Time Delay Integration (TDI) image sensors. High performance CCD-in-CMOS devices use buried channel to achieve efficient multiple charge transfer and can have high conversion gain and low noise on a par with the PPD, thanks to a small sense node. BC process modules are now routinely offered by several CIS foundries to help the design of TDI imagers. Buried channel CCD-in-CMOS devices have demonstrated charge transfer inefficiency (CTI) as low as $10^{-5}$ per transfer [18]. However, their dark current is much higher than in PPDs because the Si-SiO$_2$ interface at the BC is not inverted.

Based on these considerations, we have chosen the PPD as the photosensitive element in this development due to its superior dark current, combined with a BC CCD-based structure for the multiple signal sampling. Figure 1 shows the simplified diagram of the proposed pixel. As done normally, the charge is transferred from the PPD by the transfer gate (TG). Following the TG is a BC structure, where the charge is kept, transferred and sampled multiple times non-destructively.

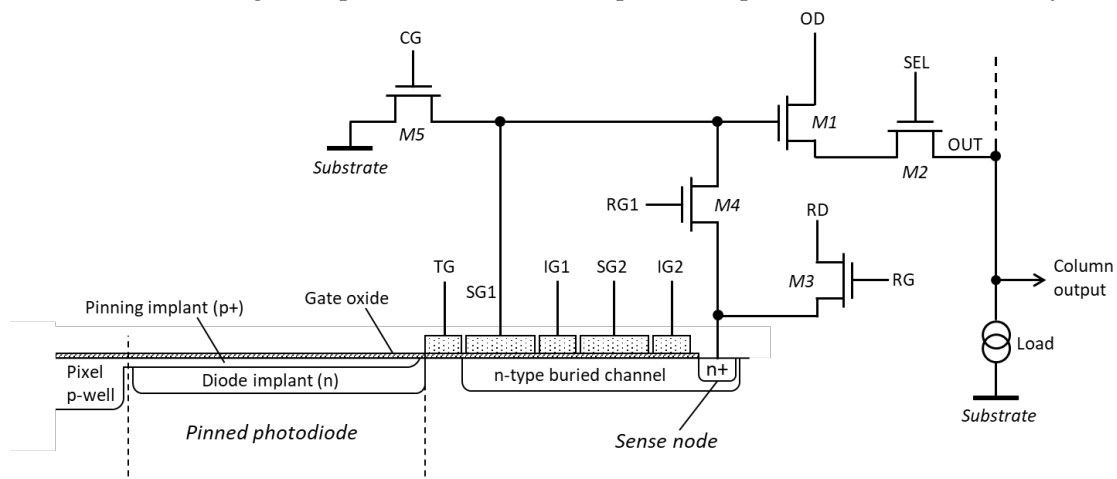

Figure 1. Diagram of the proposed pixel.

The first sense gate (SG1) is used to capacitively measure the signal stored under it. The voltage on SG1 is buffered by the source follower M1, which connects to the column output line via the row select transistor M2. Here OD (output drain) is the DC supply to the source follower and SEL is the control signal to the gate of the row select transistor. The transistor M5 is used to lower the potential on SG1 to substrate so that the charge can be transferred to SG2.

Figure 2 shows the potential diagrams of the pixel during its operation. Before the charge is transferred out of the PPD, in step 1 the voltage on SG1 is reset to the reset drain (RD) potential by turning both transistors M3 and M4 on. In step 2 the transfer gate (TG) is pulsed high and the charge stored under the PPD is transferred under SG1, followed by the turn-off of TG in step 3. In step 4 the storage gate SG2 is reset and the insulating gate IG1 is biased in preparation for the charge transfer from SG1 to SG2 in step 5. The gate SG1 is lowered to substrate potential by applying a pulse to the gate CG of M5 while M3 and M4 are turned off, which makes the charge move under the gate SG2.

Once the charge has been transferred as shown in step 6, SG1 is reset by turning on M3 and M4 in step 7, with the M5 off. After this the transistors M3 and M4 are turned off and the reset level on SG1 is read out via the source follower M1 as required to implement correlated double sampling (CDS). Next, the potential on SG2 is lowered in step 8 so that the charge can be transferred back under SG1. After the transfer is complete, the amount of charge under SG1 is measured by differencing the voltage level on SG1 and the previously taken reset sample. By repeating steps 4 to 10 it is possible to measure the signal charge under SG1 multiple times non-destructively.

Figure 1 shows SG1 as the floating gate used for the measurement, however SG1 and SG2 are interchangeable and either can be used if they connect to M1. Furthermore, SG1 and SG2 can be connected to their own separate readout circuits, consisting of a source follower, row select and reset transistors. Figure 2 shows that the reset and the signal time periods on SG1 and SG2 are in anti-phase, so that the reset level on SG1 can be measured simultaneously with the signal level on SG2, and vice versa, as in steps 7 and 10. This makes it possible to use both SG1 and SG2 for multiple signal sampling, thus reaching the required number of samples in half the time. The downsides of this approach are the increased transistor count which occupy larger area, and the need to have two outputs per pixel, because the signals from SG1 and SG2 appear concurrently. In addition, the gain difference between the two non-destructive readout paths would make the signal processing more challenging. In this work we are considering sampling only on SG1, as shown in Figure 1.

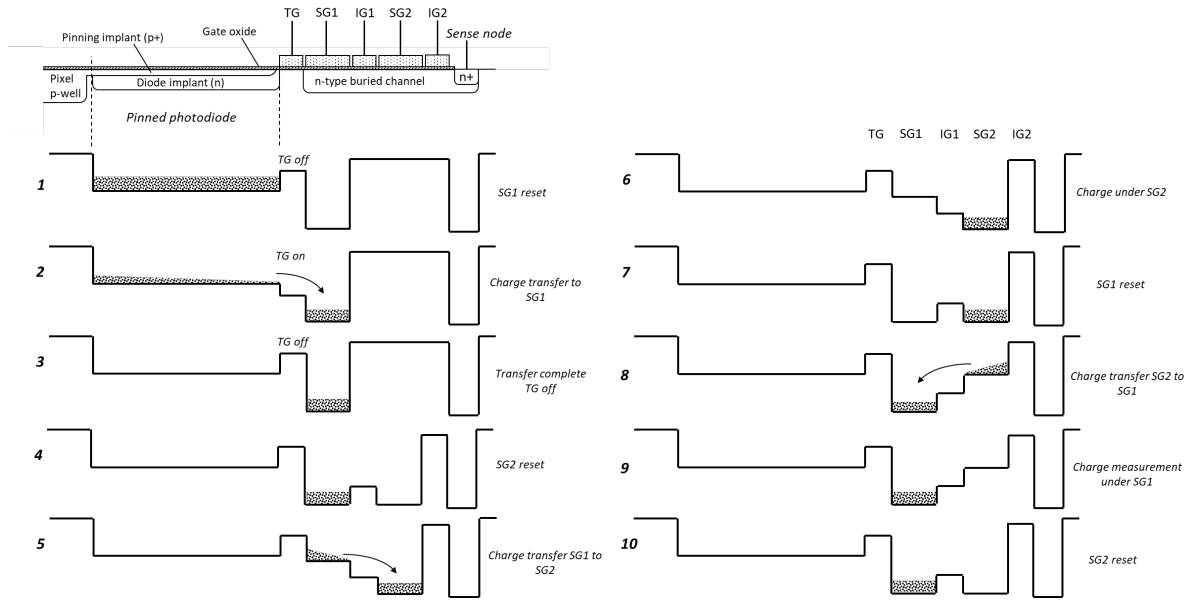

Figure 2. Timing diagram of the charge transfer in the proposed pixel.

After enough measurements have been made, the charge can be transferred out of the BC. This can be done by pulsing the gate IG2 after the sense node has been reset in advance by the transistor M3, while M4 is on. This allows conventional readout of the signal with CDS, after which the charge is destroyed. Alternatively, the charge can simply be dumped to the sense node without reading it. In any case, the charge in the BC must be cleared before the next charge is transferred from the PPD.

The pixel architecture relies on efficient multiple charge transfer in a BC CCD with low channel potential $V_{ch0}$. Typically, the PPD pinning voltage is in the range between 1.0 and 1.5 V, which implies that the CCD channel potential must be of similar value. We have performed TCAD simulations based on 180 nm CIS process using a customized BC implant to achieve $V_{ch0} = 2$ V. The simulations model a PPD and a BC with 4 gates as shown in Figure 2, using Athena (for process and device generation) and Atlas (for device simulation) commercial software from Silvaco Inc. Using simulated light, charge is generated in the PPD and collected, and then transferred out to the BC by pulsing the transfer gate TG. This is not different from normal PPD operation.

Figure 3(a)-(c) shows the potential in the BC and the size of the stored charge during transfer, demonstrating that the charge is stored away from the Si-SiO$_2$ interface, which is needed for good CTI. Figure 3(d) shows the potential along the middle of the buried channel when IG1 is biased to 0.5 V and acts as a potential barrier between SG1 and SG2, both biased at 2 V. During the charge transfer between SG1 and SG2 the IG1 gate is held at an intermediate constant voltage in order to reduce the capacitive coupling to SG1 from the clock voltage on SG2. Simulations indicate that the charge transfer between SG1 and SG2 completes within 0.2 ns for signals of less than 100 electrons, which is expected considering the short gate length. Because of that, the charge transfer time is likely to be negligible compared to the signal sample time, which is in the range of few hundred nanoseconds.

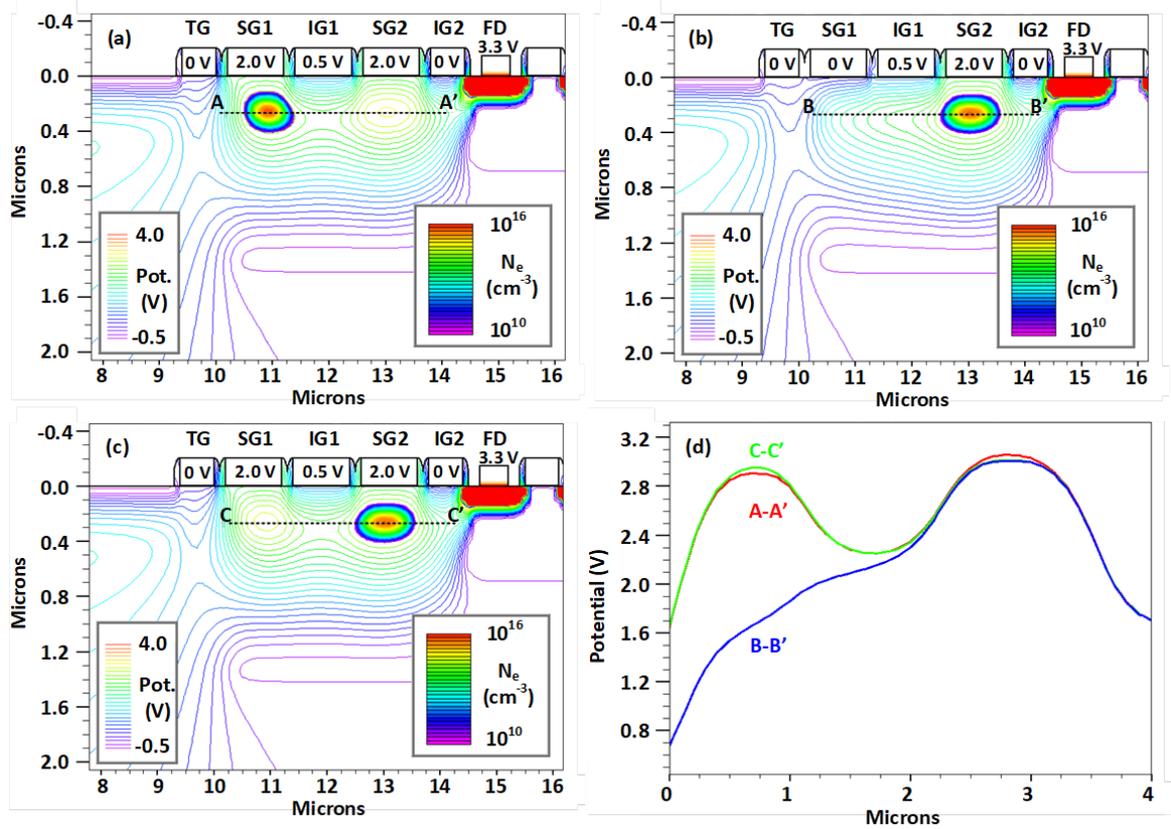

Figure 3. Simulated potential profile in the buried channel and electron density plots for charge stored under SG1 (a), during transfer from SG1 to SG2 (b) and for charge stored under SG2 (c). Plot (d) shows the potential along the dashed lines in (a), (b) and (c).

## 3. Expected performance

### 3.1. Readout noise

We can calculate the expected noise performance of a CIS using the described multiple non-destructive signal sampling. The following considerations assume that a whole pixel row is read out simultaneously using column-parallel circuitry, as is the norm in most CMOS image sensors.

The voltage noise density of the source follower $e_n$ (in units of V/$\sqrt{Hz}$) as a function of the frequency $f$ can be expressed as

$$e_n = e_{nw}\sqrt{1 + f_{nc}/f}, \quad (1)$$

where $e_{nw}$ is the white noise density and $f_{nc}$ is the $1/f$ noise corner frequency, defined as the frequency where the white and the 1/f noise have the same power. The signal readout is assumed to use dual slope integrator (DSI) circuitry, which has nearly ideal Signal-to-Noise Ratio (SNR) for signals dominated by white noise and with negligible 1/f noise. The noise performance of the DSI can be approximated by digital multi-sampling [19], which is widely used in modern low noise CIS. With the in-pixel source follower being the dominant noise source, the RMS noise voltage $V_n$ at the output of a DSI circuit is given by [20]

$$V_n = e_{nw}\sqrt{2f_r + 4f_{nc}\ln 2}. \quad (2)$$

where $f_r$ is the readout frequency. The time period of one signal measurement is $T_r = 1/f_r$ and comprises the reset and the signal samples. To convert to input-referred ENC we divide (2) by the charge-to-voltage conversion gain $S_V = q/C_n$, where $q$ is the elementary charge and $C_n$ is the total capacitance of the sense node. The ENC of one signal measurement becomes

$$Q_{n\_1} = \frac{e_{nw}}{S_V}\sqrt{2f_r + 4f_{nc}\ln 2}. \quad (3)$$

After performing $N$ independent signal measurements the ENC of the average decreases to $Q_{n\_N} = Q_{n\_1}/\sqrt{N}$. Therefore, to reach the ENC threshold for single electron detection $Q_{n\_sp}$, taken as $Q_{n\_sp} = 0.15$ e- RMS [5], the number of measurements $N_{sp}$ must be

$$N_{sp} = (Q_{n\_1}/Q_{n\_sp})^2. \qquad (4)$$

Substituting (3) into (4) gives the number of the required signal measurements

$$N_{sp} = \left(\frac{e_{nw}}{S_V Q_{n\_sp}}\right)^2 (2f_r + 4f_{nc}\ln 2). \qquad (5)$$

The time to read one pixel, equal to the readout time of one image row $t_{row}$, becomes

$$t_{row} = N_{sp} T_r = \left(\frac{e_{nw}}{S_V Q_{n\_sp}}\right)^2 \left(2 + \frac{4f_{nc}\ln 2}{f_r}\right). \qquad (6)$$

Whether an imager using multiple signal averaging can find practical application depends on the number of samples $N_{sp}$ needed to reach single-photon sensitivity. From (5) we can see that to reduce $N_{sp}$ it is very important to reduce $e_{nw}$ and maximize $S_V$ due to the quadratic dependence. At readout frequencies well above the noise corner frequency, which are likely to be used in order to reduce the readout time, the second term in (6) levels off to two.

After optimization of the source follower's channel length and width, its noise performance is determined mainly by the foundry process, leaving the conversion gain $S_V$ as the most important parameter in (6) to address. Using typical transistor noise parameters for 180 nm CMOS image sensor process, we can calculate the number of samples and the row readout time as a function of $S_V$. As a representative example, Figure 4 shows the number of samples $N_{sp}$ required to achieve single-photon sensitivity ($Q_{n\_sp} = 0.15$ e- RMS), calculated from (5) for $e_{nw} = 30$ nV/$\sqrt{\text{Hz}}$ and $f_{nc} = 330$ kHz. These parameters are typical and are not taken for a specific foundry, however we expect to see gradual improvements in the future from advances in the processing technology for low noise transistors.

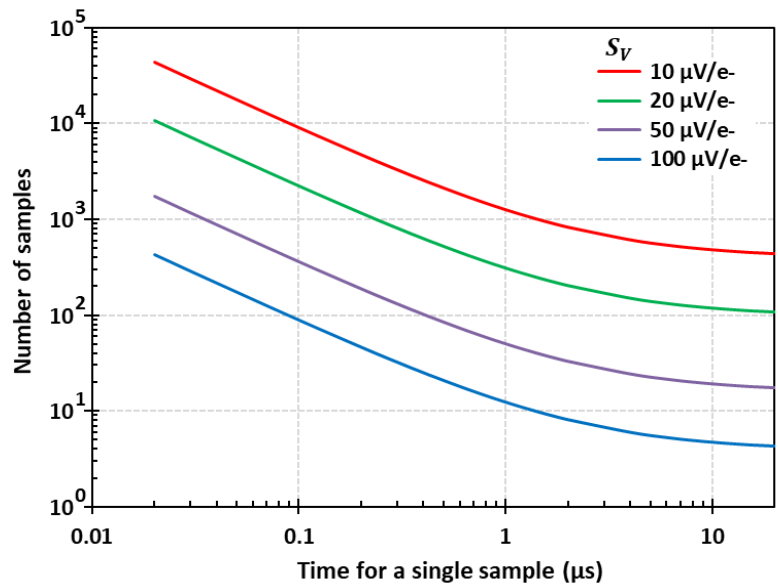

Figure 4. Number of averaged samples for 0.15 e- RMS ENC for different conversion gains as a function of the time required for a single sample for $e_{nw} = 30$ nV/$\sqrt{\text{Hz}}$ and $f_{nc} = 330$ kHz.

Figure 4 shows that the required number of samples falls as $1/T_r$ and levels off at long sample times due to the increasing fraction of $1/f$ noise in the total. We are interested in performing large number of samples in a short amount of time, therefore it is important that the $1/f$ noise corner frequency is well below the readout frequency.

Figure 5 shows the number of image sensor rows that can be readout per second, equal to $1/t_{row}$ calculated from (6) for the same conditions as in Figure 4. For the highest used conversion gain ($S_V = 100$ μV/e-) we can see that a megapixel-scale sensor with 1000 rows and >1000 columns can achieve readout rate exceeding 100 fps when the time for one sample is shorter than about 300 ns. This is within the capabilities of existing CIS technology and indicates that this approach can be practical. Even for $S_V$ = 10 µV/e-, which should be easier to achieve with a floating gate, a megapixel-scale sensor should be able to reach one frame per second.

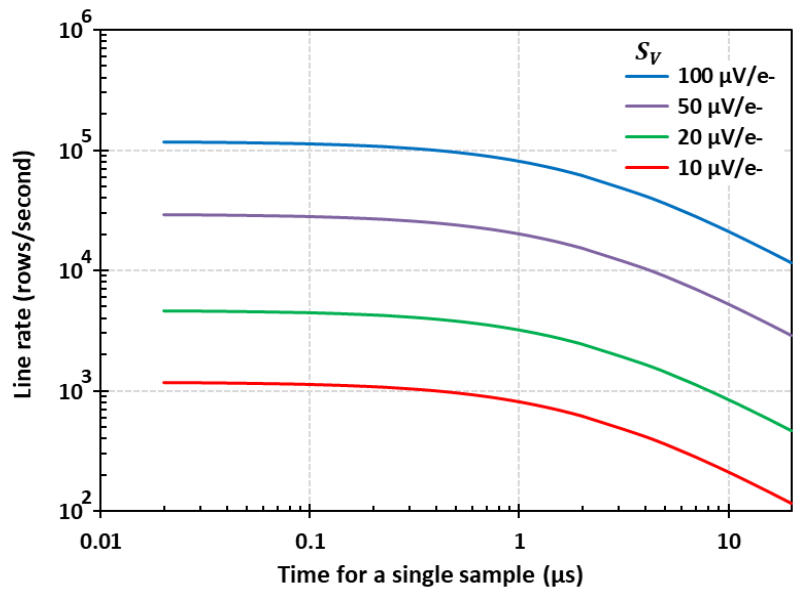

Figure 5. Line rate for 0.15 e- RMS ENC for different conversion gains as a function of the time required for a single sample.

The readout rates calculated above depend very much on the noise characteristics of the used CIS process and whether the required conversion gain can be achieved. Further on, we investigate pixel layouts based on the TCAD simulations shown in Figure 3 and on the design rules of a 180 nm process from a leading CIS foundry.

*3.2. Linearity and full well capacity*

It is important that the method of capacitive charge measurement is linear, i.e. the signal depends linearly on the amount of charge under SG1. A simulation of the voltage difference between the reset and the signal samples at SG1 was performed for the model in Figure 3 at increasing signal sizes, and the result is shown in Figure 6. The signal response has linearity, defined as the deviation of the data from the line of best fit, of better than 1 %. The slope of the straight-line fit gives a conversion gain of 83.4 µV/e- for a SG1 gate with length of 1 µm. The effective gate capacitance of SG1 in this model does not include contributions from the source follower and the reset transistor in Figure 1, but takes into account the gate capacitance to the BC and the gate-to-gate edge capacitances.

The depth of the 2D model is 1 µm, which corresponds to a BC width of the same size. In the simulation the charge stored under SG1 begins to fill up the potential well and gets in contact with the Si-SiO$_2$ interface above 2000 e-, which severely degrades the CTI. This corresponds to a full well capacity (FWC) of 2 ke- per µm of channel width. If needed, the signal capacity can be increased by widening the BC at the expense of reducing the conversion gain, since SG1 is widened as well.

The FWC of the proposed sensor is dominated by the BC CCD due to its small size, and the area taken up by the PPD is also smaller than in a typical PPD pixel. This would not be a major limitation if high readout rate is used due to the ability to stack the images without additional noise.

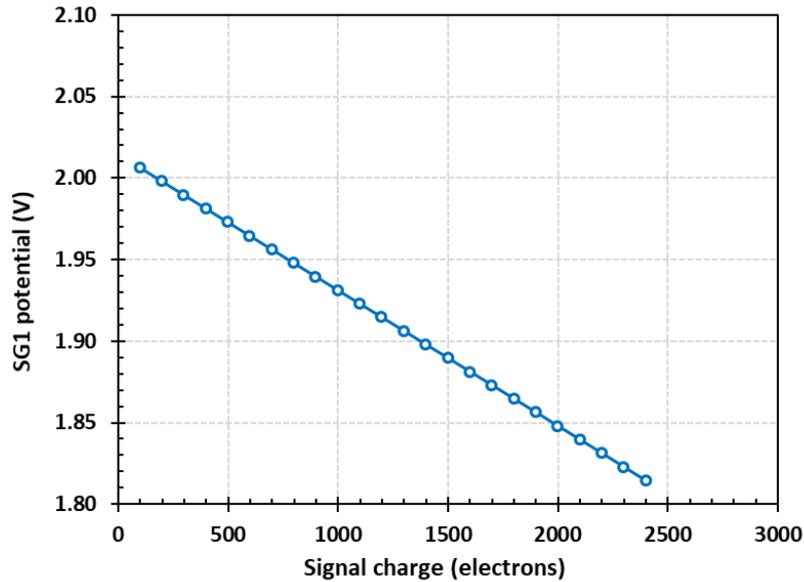

Figure 6. Simulated voltage on SG1 as a function of the amount of charge under SG1. The simulation is in 2D with 1 µm depth in the third dimension, corresponding to the BC channel width.

## 4. Design

### 4.1. Pixel design

Based on a reference 10 µm pitch PPD pixel, we have created several designs suitable for implementation in a prototype image sensor. Figure 7 shows the bottom part of the PPD, the buried channel and the readout transistors of an example pixel layout. The PPD occupies approximately 40% of the pixel area, and the BC extends from the TG in a straight line towards the edge of the pixel. More compact layouts with higher fill factors are possible if the BC is bent by 90 degrees under the PPD.

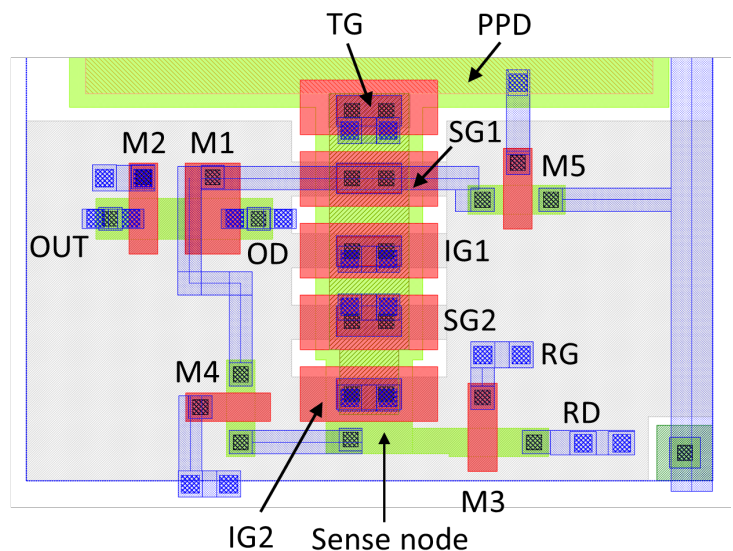

Figure 7. Example pixel layout using 180 nm CIS process, showing the BC elements and the readout circuitry. Only the bottom part of the PPD is displayed.

In this layout the total capacitance of SG1 has been estimated by capacitance extraction from the Cadence Virtuoso CAD software, giving 2.63 fF and $S_V$ = 61 µV/e-. This conversion gain is lower than the one derived from the TCAD simulation in Section 3.2 due to the inclusion of the source follower and all other parasitic capacitances, but is still high enough to allow sensors with single-photon sensitivity to be built. Using 1 MHz readout frequency, equation (3) gives ENC = 0.88 e- RMS for a single measurement, and a line rate of nearly 30000 rows/s with averaging of 34 samples to finally achieve ENC = 0.15 e- RMS. This would translate to 30 fps from a 1 Mpixel sensor. In comparison, an EMCCD with the same format (e.g. CCD201-20 from Teledyne e2v) achieves 14 fps with sub-electron noise at 15 MHz serial pixel rate.

*4.2. Performance limitations*

The suitability of the proposed sensor for its target applications depends on many factors. The first and perhaps the most important one is the noise for a single charge measurement, which we addressed in Section 3. The other factors we need to consider are the image lag in the PPD, the CTI in the buried channel and the dark current in the device.

The PPD can suffer from incomplete charge transfer, called image lag [16], due to the slow electron transport and potential barriers and pockets along the transport path. Various methods involving potential gradients inside or outside the PPD have been developed to reduce the image lag. The most widely used one creates a potential gradient under the transfer gate using an implant and achieves image lag below 0.1% [17]. At this low level the lag is unlikely to be a limiting factor.

Tens or even hundreds of charge transfers have to be made during the multiple signal measurement, and the charge can spend long time in the BC, during which it can be captured by traps. If the trap emission time constant is short, some of the charge can be re-emitted into its own signal packet, otherwise it will merge with the next one and will be lost. We can expect the trap concentration in modern CIS processes to be low, and its impact on the CTI can be estimated from the available literature data.

The worst-case CTI is seen when every trap in the volume swept by the charge packet is empty and able to trap an electron, a process characterized by a capture time constant [21]. If subsequent re-emission into the same packet does not occur the CTI is simply

$$CTI = \frac{n_t}{n_s}, \qquad (7)$$

where $n_t$ is the trap density and $n_s$ is the signal charge density. To our knowledge, the best CTI achieved in a BC CCD-in-CMOS device is $10^{-5}$ for a 5 μm pixel [18]. The usual signal size for a CTI measurement is 1620 electrons, generated by X-rays from a $^{55}$Fe source. From (7) we can calculate that a CTI of $10^{-5}$ corresponds to 0.0162 active traps in a pixel area of 5 μm × 5 μm, since the traps and the signal occupy the same volume. The channel area used for charge transfer in the layout in Figure 7 is approximately 2 μm × 3 μm; this translates to $3.9×10^{-3}$ traps/pixel, or one in 257 pixels containing a trap. Fortunately, the capture time constant for small charges is large [21] because it is inversely proportional to the signal density. Therefore, we can expect the effects of charge trapping to be much less than the worst case described above, which assumes that every trap captures an electron.

The buried channel operates in non-inverted mode and therefore its dark current density is likely to be in the region of several nA/cm$^2$ [18] at room temperature. If we take the optimistic value of 1 nA/cm$^2$, counting on future technology improvements, the dark current in the BC with an area of 6 μm$^2$ is 375 e-/s. During a typical row readout time of 30 μs (corresponding to 30 samples at 1 μs each) this would generate 0.011 dark current electrons on average. Using a metal shield over the buried channel in front-side illuminated sensors may be necessary to block out parasitic light and prevent spurious signals. In back-side illuminated sensors a buried p-well under the BC can be used instead.

The dark current density from the PPD is much lower due to the pinning implant and is usually below 1 pA/cm$^2$, however the PPD collects charge from the whole pixel area. If we take the dark current density in the PPD as 1 pA/cm$^2$ and the pixel area as 10 μm × 10 μm, the number of dark current electrons per pixel at 100 fps will be 0.0625. Longer integration times can increase this figure to a level where single-photon detection becomes compromised.

Dark current shot noise from the BC and the PPD adds in quadrature with the electronic noise and is a fundamental limitation to single photon imaging. Short exposure times and cooling can be used to reduce the dark current. For long exposures the operating temperature of the sensor has to be reduced significantly, and can be as low as -100 °C, as it typical for astronomical CCDs.

*4.3 Sensor architecture*

Figure 8 shows the block diagram of the proposed SP image sensor. Similar to a mainstream CIS the sensor includes a row address decoder, row clock drivers, CDS circuitry and an ADC for each pixel column. The raw data rate from the ADC array is likely to be very high due to the multiple signal samples, therefore it would be much more convenient and power efficient to perform the signal averaging on chip. Taking data from the ADCs, a dedicated circuitry in each column is added to perform the digital accumulation of the multiple samples, which is then scaled to obtain the signal average. A dual slope ADC which integrates the reset and the signal levels with opposite signs could also perform the role of an averager, when operating on each sample consecutively. Following the digital averager the data is multiplexed and read out serially on one or several sensor outputs. In the block diagram in Figure 8 the row and the column address and the pixel control signals are provided externally to the image sensor, however they can be generated on chip by a state machine residing in the control block.

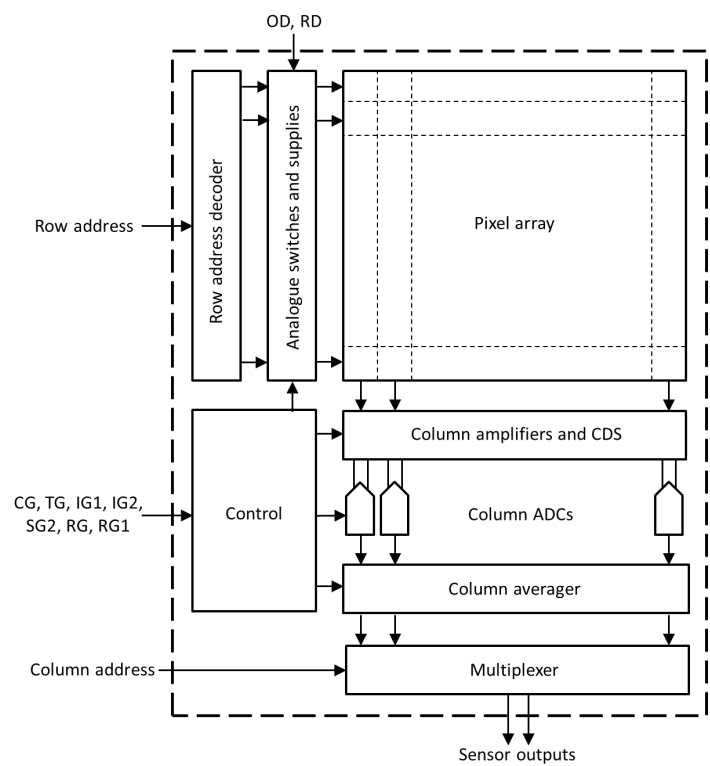

Figure 8. Block diagram of the proposed image sensor.

The increase of peripheral circuitry compared to a traditional CIS is modest because accumulation and scaling are relatively simple operations which can be implemented efficiently in digital logic. In a more advanced design using 3-D integration [22] the peripheral circuitry including the averaging can be placed on a separate digital tier.

The expected performance of the proposed image sensor can be compared with existing devices such as skipper CCDs, EMCCDs, high conversion gain CIS and SPADs. A major advantage of the proposed implementation in CMOS technology over the skipper CCD [13] is the greatly increased frame rate, which could be three orders of magnitude higher. This is due to the massive parallel readout possible in CIS, while offering the same or better noise performance. This sensor could have similar characteristics to an EMCCD operated in a photon counting mode [4] but without the drawbacks of using high voltages, such as gain ageing and clock-induced charge generation, which reduces the photon detection probability [23]. CIS with high conversion gain do not offer true single photon sensitivity yet [7]. Our proposed design has adjustable noise characteristics depending on the number of samples and can be better than high conversion gain CIS, but is slower due to the multiple sampling. Our design is also much slower than SPADs [3], but does not suffer from high dark count rate due to the absence of avalanche gain. This skipper CMOS design therefore provides several advantages over existing technologies, combining single-photon detection at high frame rates with CMOS compatible-voltages and no spurious signal.

## 5. Conclusions and outlook

We have presented a concept for a single-photon image sensor using multiple non-destructive signal sampling in CMOS image sensor technology. By averaging multiple signal samples, which are statistically independent, the readout electronic noise can be reduced by the square root of the number of samples. This allows the noise level to be reduced to ENC < 0.15 e- RMS, a level widely considered necessary for single electrons to be reliably distinguished. This will allow imaging limited only by the shot noise of the registered photons, and free from electronic noise.

The proposed pixel design uses a pinned photodiode as a photosensitive element and a buried channel CCD per pixel for multiple charge transfer and non-destructive signal readout by capacitive coupling between a sense gate and the signal charge. A pixel design using this method has been simulated in TCAD and several layouts have been generated for a 180 nm CMOS image sensor process. Significant increase in the readout speed over the equivalent "skipper CCD" technique is possible due to the massive parallel readout in CMOS image sensors. Using simulations, the noise performance of the pixel has been determined as a function of the number of samples, sense node capacitance, sampling rate, and transistor characteristics.

The presented results show that single-photon imaging using multiple non-destructive signal sampling in CIS technology is viable. Our designs and simulations indicate that a megapixel-scale sensor operating at ~100 fps is feasible using present-day technology. Such a sensor could find numerous applications in science and technology in fields such as astronomy, adaptive optics, biological imaging, quantum technologies, and autonomous systems.

Future work will involve the manufacturing of a prototype CIS using several variants of the described pixel design. Following that, a full-scale imager with digital signal averaging implemented on chip will be the next objective of our work.

**Acknowledgments:** This project has received funding from the ATTRACT project funded by the EC under Grant Agreement 777222.

**Author Contributions:** K.D.S. led the project, performed TCAD simulations and calculations of the expected performance and wrote the paper; M.P. performed TCAD simulations, schematic designs and simulations, and laid out the pixel designs in CAD; M.D., E.G. and N.B. contributed with design ideas, paper writing, reviewing and editing; A.D.H. contributed to conceptualization, project management and academic supervision.

**Conflicts of Interest:** The authors declare no conflict of interest.